\documentclass[prl, showpacs, amsmath,amssymb, byrevtex,onecolumn]{revtex4}
\date{}
\usepackage{graphicx}
\usepackage{color}
\usepackage{dcolumn}
\usepackage{amsmath}
\usepackage{epsfig}

\begin{document}
\title{Deterministic Quantum Key Distribution Using Gaussian-Modulated Squeezed States}
\author{Guangqiang He}
\email{gqhe@sjtu.edu.cn}
%\author{Jun Zhu}
%\email{bierhoff_24@126.com}
\author{Guihua Zeng}
\email{ghzeng@sjtu.edu.cn}
%\author{Jin Xiong}
%\email{xiongjint@sjtu.edu.cn}
\affiliation{National Key Laboratory on Advanced Optical Communication Systems and Networks, \\
Department of Electronic Engineering, Shanghai Jiaotong University, Shanghai 200030,China}
\date{\today}
%\maketitle

\begin{abstract}
A continuous variable ping-pong scheme, which is utilized to
generate deterministically private key, is proposed. The proposed
scheme is implemented physically by using Gaussian-modulated
squeezed states. The deterministic way, i.e., no basis
reconciliation between two parties, leads a two-times efficiency
comparing to the standard quantum key distribution schemes.
Especially, the separate control mode does not need in the
proposed scheme so that it is simpler and more available than
previous ping-pong schemes. The attacker may be detected easily
through the fidelity of the transmitted signal, and may not be
successful in the beam splitter attack strategy.

%\vspace{0.8cm}
\end{abstract}

\maketitle

The standard quantum key distribution (QKD) scheme
\cite{Gisin2002} provides a novel way of generation and
distribution of secret key. Its security is guaranteed by the law
of quantum mechanics \cite{Lo1999,Shor2000,Mayers2001}. The
intrinsical basis reconciliation, which is significant in
guaranteeing the security, means that the standard QKD is
nondeterministic. Unfortunately, the nondeterministic property
results in loss of many qubits, consequently, the efficiency is
very low. To improve the efficiency, several deterministic QKD
schemes were proposed recently by using technique of ping-pong of
photon between two parties
\cite{Bostrom2002,QingYu2003,Lucamarini2005}. These schemes are
implemented in discrete variable. While the final key is generated
through a message-mode and the security is guaranteed by a
separate control-mode. However, the separate control-mode in the
previous ping-pong schemes leads a higher communication complexity
and a more complicated experimental realization. In addition,
discrete variable is not easy in generation as well as detection
so that continuous variable (CV) becomes a favored candidate in
the quantum cryptography \cite{Braunstein2005,
Ralph1999,Hilery2000,Gottesman2001,Cerf2001,Grosshans2002,
Silberhorn2002,Grosshans2003}.

In this letter, a continuous variable ping-pong scheme, which is
implemented by using the Gaussian-modulated squeezed states, is
firstly proposed. Since the proposed scheme does not need the
basis reconciliation when the communicators, i.e., Alice and Bob,
exchange the key information, its efficiency is two times of the
standard CV QKD schemes. Particularly, the separate control-mode,
which is necessary in the discrete-variable (DV) ping-pong
schemes, can be omitted. This characteristic makes the proposed
scheme be feasible in experimental realization. In addition, the
channel capacity is higher than that of the DV ping-pong schemes.
The security analysis based on Shannon information theory shows
clearly the security against the beam splitter attack strategy. In
a lossy channel, when the transmission is larger than $0.728$ the
security can be warranted.

\begin{figure}[htp1]
\begin{center}
\includegraphics [width=70mm,height=45mm]{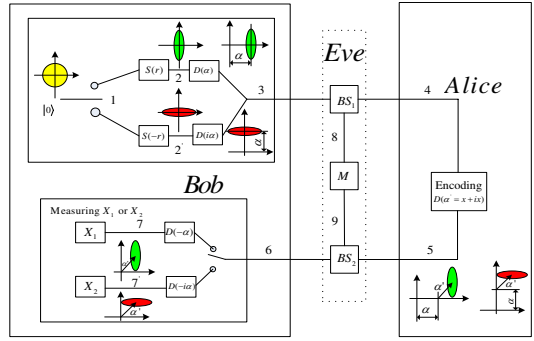}
\end{center}
\caption{Scheme of the deterministic quantum key distribution
using Gaussian-modulated squeezed states. The Arabian numbers
denote the modes in Heisenburg picture. } \label{protocol}
\end{figure}

The proposed scheme, which is sketched in Fig.\ref{protocol},
executes the following. Step 1, Bob operates an initial vacuum
state $|0\rangle$ with either operator ${\mathcal
P}=D(\alpha)S(r)$ or ${\mathcal P}_\bot=D(i\alpha)S(-r)$ which are
regarded as a pair of bases, where $S(r)$ is the squeezing
operator and $r$ is the squeezed factor, $D(\alpha)$ is a
displacement operator which is employed to add noise, and $\alpha$
is a real number. Step 2, Bob sends the generated state to Alice
through a public quantum channel. Step 3, having operated the
received state by a proper displacement operator
$D(\alpha'=x+ix)$, Alice returns the state to Bob, where $x$ is a
random number which is drawn from Gaussian probability
distribution. Step 4, if the base $\mathcal P$ is employed in the
step 1, Bob applies $D(-\alpha)$ on the returned mode, and
measures $X_1$. Otherwise, Bob applies $D(-i\alpha)$ on the
returned mode, and measures $X_2$. The canonical quadratures $X_1$
and $X_2$ are defined as $X_1=\frac{1}{2}(\hat{a}+\hat{a}^{\dag})$
and $X_2=\frac{1}{2i}(\hat{a}-\hat{a}^{\dag})$. Step 5, Alice
randomly selects some $x$ values, and then sends the chosen values
and their corresponding time slots to Bob through a public classic
channel. Step 6, after has received Alice's values, Bob calculates
statistically the fidelity by using the received values and his
corresponding measured results. Then Bob detects whether Eve is
absent or not by using the calculated fidelity.

We explain briefly above protocol in the physical way. Bob's operation in step 1 yields a squeezed
coherent state either $|\alpha,r\rangle= D(\alpha)S(r)|0\rangle$ or $|i\alpha,-r\rangle=
D(i\alpha)S(-r)|0\rangle$. For simplicity, we only illuminate the evolvement of the state
$|\alpha,r\rangle$ thereafter since the state $|i\alpha,-r\rangle$ may be treated with in a same
way. Since the initial vacuum state $|0\rangle$ is a Gaussian state, the canonical quadratures
$X^1_1$ and $X^1_2$ follow the Gaussian probability distribution, i.e., $X_1^1\sim
N(0,\frac{1}{4})$ and $X_2^1\sim N(0,\frac{1}{4})$, where $X^{i}_{k} (k=1, 2)$ represents $X_{k}
(k=1, 2)$ of the mode $\hat{a}_{i}$, $\Gamma\sim N(\mu, \sigma^2)$ denotes that random variable
$\Gamma$ follows Gaussian probability distribution with the average value $\mu$ and the variance
$\sigma^2$. Choosing a random disturbance $\alpha$ with distribution $A\sim N(0,\Sigma^2)$, one has
$X_1^3\sim N(0,\Sigma^2+\frac{1}{4}e^{-2r})$ and $X_2^3\sim N(0,\frac{1}{4}e^{2r})$. Obviously,
when the following condition is satisfied, i.e.,
\begin{equation}\label{condition}
\Sigma^2+\frac{1}{4}e^{-2r}=\frac{1}{4}e^{2r},
\end{equation}
$X_1^3$ and $X_2^3$ follows the same probability distribution. Subsequently, Eve cannot distinguish
the output states $|\alpha,r\rangle$ and $|i\alpha,-r\rangle$ whatever the statistics Eve
accumulates. Making use of the operator $D(\alpha'=x+ix)$ and the distribution $X\sim
N(0,\Sigma{'}^2)$, one may easily obtain $X_1^{5}\sim
N(0,\Sigma^2+\frac{1}{4}e^{-2r}+{\Sigma^{'}}^2)$ and $X_2^{5}\sim N(0,\frac{1}{4}e^{2r}+{\Sigma
^{'}}^2)$. After has received the state encoded by Alice, Bob removes the added quantum noise so
that he can decode Alice's message. This operation gives the mode $\hat{a}_{7}$ with $X_{1}^7\sim
N(0,\frac{1}{4}e^{-2r}+{\Sigma^{'}}^2)$ and $X_{2}^7\sim N(0,\frac{1}{4}e^{2r}+{\Sigma^{'}}^2)$.
Finally, Bob measures $X_{1}^7$ on the received state to decode Alice's message.

Now we move on the security analysis. Suppose Eve splits the forward and backward beams as depicted
in Fig.\ref{protocol}, then she coherently measures the intercepted beams to obtain the maximal
information. To show the security of the proposed scheme against above attack strategy, i.e., the
general beam splitter attack strategy,  we adopt the following criterion which is used prevalently
for QKD scheme \cite{Gisin2002,Maurer1993},
\begin{equation}
\Delta I=I(\alpha,\beta)-I_{max}(\alpha,\epsilon)>0,
\end{equation}
where $I(\alpha,\beta)$ is the mutual information between Alice and Bob, and
$I_{max}(\alpha,\epsilon)$ is the maximal mutual information between Alice and Eve. According to
Shannon information theory \cite{Shannon1948}, the channel capacity of the additive white Gaussian
noise (AWGN) channel is given by,
\begin{equation}\label{shannon}
I=\frac{1}{2}\log_2(1+\gamma),
\end{equation}
where $\gamma=P_S/P_N$ is the signal-noise ratio, $P_S$ and $P_N$ are the variances of the signal
and noise probability distributions respectively. If the signal follows the Gaussian distribution,
and the channel is an AWGN channel, the channel capacity is the mutual information of the
communication parties. In the followings, first we calculate the probability distributions of $X_1$
and $X_2$ in all modes as depicted in Fig.1, then calculate $I(\alpha,\beta)$ and
$I_{max}(\alpha,\epsilon)$ according to Eq.(\ref{shannon}).

A beam splitter (BS) is always employed to split the laser beam.
According to Fig.\ref{protocol}, inputs of the $BS_1$ are given
by,
\begin{eqnarray}\label{BS1in}
X_1^{3}=e^{-r}X_1^1+A, \,\,\, X_2^{3}=e^{r}X_2^1.
\end{eqnarray}
and two output modes of $BS_1$ are,
\begin{eqnarray}\label{BS1out}
X_1^4=\sqrt{\eta_1}X_1^3+\sqrt{1-\eta_1}X_1^{vac1}, \nonumber \\
X_2^4=\sqrt{\eta_1}X_2^3+\sqrt{1-\eta_1}X_2^{vac1}, \nonumber \\
X_1^8=\sqrt{\eta_1}X_1^{vac1}-\sqrt{1-\eta_1}X_1^3, \nonumber \\
X_2^8=\sqrt{\eta_1}X_2^{vac1}-\sqrt{1-\eta_1}X_2^3,
\end{eqnarray}
where ${\eta_1}$ is the transmittance coefficient of $BS_1$. Consider Alice's operation of applying
the displacement operator $D(\alpha^{'}=x+ix)$ on the mode $\hat{a}_4$, the inputs of the $BS_2$
are given by,
\begin{eqnarray}\label{BS2in}
X_1^{5}=X_1^4+X, \,\,\,  X_2^{5}=X_2^4+X.
\end{eqnarray}
Similarly, the outputs of the $BS_2$ are obtained as following,
\begin{eqnarray}\label{BS2out}
X_1^6=\sqrt{\eta_2}X_1^5+\sqrt{1-\eta_2}X_1^{vac2}, \nonumber \\
X_2^6=\sqrt{\eta_2}X_2^5+\sqrt{1-\eta_2}X_2^{vac2}, \nonumber \\
X_1^9=\sqrt{\eta_2}X_1^{vac2}-\sqrt{1-\eta_2}X_1^5, \nonumber \\
X_2^9=\sqrt{\eta_2}X_2^{vac2}-\sqrt{1-\eta_2}X_2^5,
\end{eqnarray}
where ${\eta_2}$ is the transmittance coefficient of $BS_2$. Applying the operator $D(-\alpha)$ on
mode $\hat{a}_6$ yields,
\begin{eqnarray}\label{X7}
X_1^{7}=X_1^6-A, \,\,\,\,\,\, X_2^{7}=X_2^6.
\end{eqnarray}
Combining Eqs.(\ref{BS1in}) $\sim$ (\ref{X7}) gives,
\begin{eqnarray}\label{X7'}
X_1^{7}&=&\sqrt{\eta_2}[\sqrt{\eta_1}(e^{-r}X_1^1+A)+ \sqrt{1-\eta_1}X_1^{vac1}+X]+
\sqrt{1-\eta_2}X_1^{vac2}-A, \nonumber\\
X_2^{7}&=&\sqrt{\eta_2}(\sqrt{\eta_1}e^{r}X_2^1+ \sqrt{1-\eta_1}X_2^{vac1}+X)+
\sqrt{1-\eta_2}X_2^{vac2},
\end{eqnarray}
where the random variables $X_k^{j}$ ($k=1,2, j=vac1, vac2$) follow the Gaussian probability
distributions, i.e.,
\begin{eqnarray}\label{distribution}
X_k^{j}\sim N(0,\frac{1}{4}).
\end{eqnarray}
Making use of Eqs.(\ref{condition}), (\ref{X7'}) and (\ref{distribution}), the variances of
$X_1^{7}$ and $X_2^{7}$ are obtained,
\begin{eqnarray}\label{vX7}
\langle(\Delta
X_1^7)^2\rangle&=&\frac{1}{4}\eta_1\eta_2e^{-2r}+\frac{1}{4}(1-\sqrt{\eta_1\eta_2})^2(e^{2r}-e^{-2r})\notag\\
&&+\eta_2\Sigma^{'2}
+\frac{1}{4}[(1-\eta_2)+(1-\eta_1)\eta_2],\\
\langle(\Delta X_2^7)^2\rangle&=&\frac{1}{4}\eta_1\eta_2e^{2r}+\eta_2\Sigma^{'2}
+\frac{1}{4}[(1-\eta_2)+(1-\eta_1)\eta_2].\nonumber
\end{eqnarray}

Using the first expression in Eq.(\ref{vX7}) gives the signal-noise ratio,
\begin{eqnarray}\label{eq:sn}
\gamma_{\alpha\beta}=\frac{M}{N}.
\end{eqnarray}
where $M=\eta_2\Sigma^{'2}$ and
$N=\frac{1}{4}\eta_1\eta_2e^{-2r}+\frac{1}{4}(1-\sqrt{\eta_1\eta_2})^2(e^{2r}-e^{-2r})
+\frac{1}{4}[(1-\eta_2)+(1-\eta_1)\eta_2]$. Thus the mutual information between Alice and Bob is,
\begin{eqnarray}\label{IAB}
I(\alpha,\beta)=\frac{1}{2}\log_2(1+\gamma_{\alpha\beta}).
\end{eqnarray}
When $\eta_1=\eta_2=1$, above equation can be written as
$C(\alpha,\beta)=\frac{1}{2}\log_2(1+\frac{4\Sigma^{'2}}{e^{-2r}})$,
where $C(\alpha,\beta)$ is the mutual information between Alice
and Bob without eavesdropping. %The dependence of $C(\alpha,\beta)$
%on $r$ and $\Sigma^{'2}$ is illustrated in Fig.\ref{fig3}.
Obviously, $C(\alpha,\beta)$ increases with $r$ and $\Sigma^{'2}$.
Actually, $C(\alpha, \beta)$ is the channel capacity of the
communication between Alice and Bob without eavesdropping. As an
example, one may easily obtain $C(\alpha, \beta)=8.6$ bits when
$r=3, \Sigma'=10$, which is apparently larger than that of the DV
quantum communication scheme.
%When $r=0$, one may obtain the
%mutual information based on Gaussian-modulated coherent state.
%\begin{figure}[htp1]
%\begin{center}
%\includegraphics [width=60mm,height=40mm]{Fig.3.eps}
%\end{center}
%\caption{$C(\alpha,\beta)$ as the function of the signal variance
%${{\Sigma}^{'}}^2$ and the squeezed factor $r$. } \label{fig3}
%\end{figure}

The maximal information $I_{max}(\alpha,\epsilon)$ may be obtained by Eve through measuring both
$\hat{a}_8$ and $\hat{a}_9$. Assume that Eve is an evil quantum physicist who is able to build all
devices that are allowed by the laws of quantum mechanics. Then Eve may build an advice to measure
$X_1^{Eve}$ and $X_2^{Eve}$ of the mode $\hat{a}_{Eve}=\hat{a}_9-k\hat{a}_8$, where $k$ is a
parameter to be optimized. Using the expressions of $X^8_k$ in Eq.(\ref{BS1out}) and $X^9_k$ in
Eq.(\ref{BS2out}), $X_1^{Eve}$ and $X_2^{Eve}$ may be easily obtained,
\begin{eqnarray}\label{X1eve}
X_1^{Eve}&=&-[\sqrt{\eta_1(1-\eta_2)}-k\sqrt{1-\eta_1}](e^{-r}X_1^1+A)-\nonumber\\
&&\sqrt{1-\eta_2}X -[\sqrt{(1-\eta_1)(1-\eta_2)} +k\sqrt{\eta_1}]X_1^{vac1}
+\sqrt{\eta_2}X_1^{vac2},\\
X_2^{Eve}&=&-[\sqrt{\eta_1(1-\eta_2)}-k\sqrt{1-\eta_1}]e^{r}X_2^1-\sqrt{1-\eta_2}X-
\nonumber\\
&&[\sqrt{(1-\eta_1)(1-\eta_2)}
+k\sqrt{\eta_1}]X_2^{vac1}+\sqrt{\eta_2}X_2^{vac2}\notag.
\end{eqnarray}
Combining Eq.(\ref{condition}) and Eq.(\ref{X1eve}), one may find that the random variables
$X_1^{Eve}$ and $X_2^{Eve}$ follow the same probability distribution. Accordingly, Eve obtains the
same signal-noise ratios in $X_1^{Eve}$ and $X_2^{Eve}$, i.e.,
%\begin{eqnarray}
$\gamma_{\alpha\epsilon X_1^{Eve}}=\gamma_{\alpha\epsilon
X_2^{Eve}}={4(1-\eta_2)\Sigma^{'2}}/{\mu+\nu}$,
%\end{eqnarray}
where $\mu=(\sqrt{\eta_1(1-\eta_2)}-k\sqrt{1-\eta_1})^2e^{2r}$ and
$\nu=(\sqrt{(1-\eta_1)(1-\eta_2)} +k\sqrt{\eta_1})^2+\eta_2$. When
$k=(e^{2r}+1)\sqrt{\eta_1(1-\eta_1)(1-\eta_2)}
/(e^{2r}(1-\eta_1)+\eta_1)$, Eve obtains the maximal mutual
information,
\begin{eqnarray}\label{IAE}
I_{max}(\alpha,\epsilon)%&=&\frac{1}{2}(I_{X_1}(\alpha,\epsilon)+I_{X_2}(\alpha,\epsilon))\notag\\
&=&I_{X_1}(\alpha,\epsilon)=I_{X_2}(\alpha,\epsilon).
\end{eqnarray}

\begin{figure}[htp1]
\begin{center}
\includegraphics [width=60mm,height=45mm]{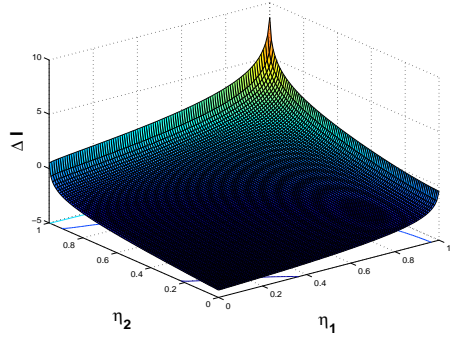}
\end{center}
\caption{Property of $\Delta I$ (unit:bit) changing with $\eta_1,
\eta_2$. The employed parameters are $r=3$ and $\Sigma^{'2}=100$.
}\label{eq:deltaI}
\end{figure}

Substituting Eqs.(\ref{IAB}) and (\ref{IAE}) into Eq.(2) gives the
secret information rate $\Delta I$. Fig.\ref{eq:deltaI} shows the
properties of $\Delta I$ changing with $\eta_1, \eta_2$. One may
see that $\Delta I$ increases with increasing of $\eta_1, \eta_2$.
In addition, $\Delta I$ may be negative when $\eta_1$ and $\eta_2$
are small, which implicates that Eve may obtain more useful
information by choosing two proper beam splitters than Bob.
Fortunately, this attack strategy does not influence the security
of the proposed scheme since Eve may be detected easily in this
situation.

\begin{figure}[htp1]
\begin{center}
\includegraphics [width=60mm,height=40mm]{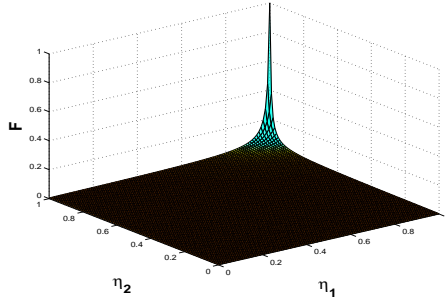}
\end{center}
\caption{The dependence of fidelity $F$ on $\eta_1$ and $\eta_2$,
the chosen parameters are $r=3$ and $\Sigma^{'2}=100$. }
\label{fidelity}
\end{figure}

Now we investigate how to detect Eve. A separate control-mode is
always employed to detect the eavesdropping in the previous
ping-pong schemes. However, this approach does not benefit the
efficiency of the scheme. Here we propose a new approach which is
more efficient than the separate control-mode approach. After
finished the step 4, Alice tells Bob some values of random
variable $X$ and the corresponding time slots through a classical
public channel. After received Alice's results, Bob calculates the
fidelity,
\begin{equation}
F=\langle\alpha_{in}|\rho_{out}|\alpha_{in}\rangle=\pi Q(\alpha_{in}),
\end{equation}
where $|\alpha_{in}\rangle=|0\rangle$, $\rho_{out}=|\alpha_{out}\rangle\langle\alpha_{out}|$, and
$|\alpha_{out}\rangle=S(-r)D(-\alpha^{'})|\Psi\rangle$ with the quantum state $|\Psi\rangle$ in
mode $\hat{a}_7$. The $Q$ function for a squeezed state is defined as that in \cite{Walls1994}. In
an ideal (no-loss) quantum channel, the fidelity satisfies $F=1$ without eavesdropping and $F<1$
with eavesdropping. Therefore the fidelity $F$ can be employed as an important parameter for Eve
detection. Making use of the state $|\alpha_{out}\rangle$ in the Heisenburg picture and
Eq.(\ref{X7'}), one obtains,
\begin{eqnarray}
X_1^{out}&=&e^r\{\sqrt{\eta_2}[\sqrt{\eta_1}(e^{-r}X_1^1+A)+ \sqrt{1-\eta_1}X_1^{vac1}+X]+
\sqrt{1-\eta_2}X_1^{vac2}-A-X\}\nonumber\\
X_2^{out}&=&e^{-r}[\sqrt{\eta_2}(\sqrt{\eta_1}e^{r}X_2^1+
\sqrt{1-\eta_1}X_2^{vac1}+X)+\sqrt{1-\eta_2}X_2^{vac2}-X].
\end{eqnarray}
Using Eqs.(10) and (17), the variances of $X_1^{out}$ and
$X_2^{out}$ are given by,
\begin{eqnarray}
\label{eq:variances} \langle(\Delta
X_1^{out})^2\rangle&=&e^{2r}\{\frac{1}{4}\eta_1\eta_2e^{-2r}+(1-\sqrt{\eta_1\eta_2})^2\Sigma^2\notag\\
&+&(\sqrt{\eta_2}-1)^2\Sigma^{'2}+\frac{1}{4}[(1-\eta_2)+(1-\eta_1)\eta_2]\}\notag,\\
\langle(\Delta
X_2^{out})^2\rangle&=&e^{-2r}\{\frac{1}{4}\eta_1\eta_2e^{2r}+(\sqrt\eta_2-1)^2\Sigma^{'2}
\nonumber\\
&+&\frac{1}{4}[(1-\eta_2)+(1-\eta_1)\eta_2]\}.
\end{eqnarray}
The fidelity $F$ is obtained as the following form,
\begin{equation}
F=\frac{2}{\sqrt{(4\langle(\Delta X_1^{out})^2\rangle+1)(4\langle(\Delta X_2^{out})^2\rangle+1)}}.
\end{equation}
Substituting Eq.(\ref{eq:variances}) into Eq.(19), one may
calculate the fidelity. Numerical solutions of Eq.(19) are
depicted in Fig.\ref{fidelity}. With $\eta_1$ and $\eta_2$
decreasing the fidelity $F$ decreases rapidly. Accordingly, any
eavesdropping can be detected by Alice and Bob by using the
fidelity $F$.

\begin{figure}[htp2]
\begin{center}
\includegraphics [width=60mm,height=40mm]{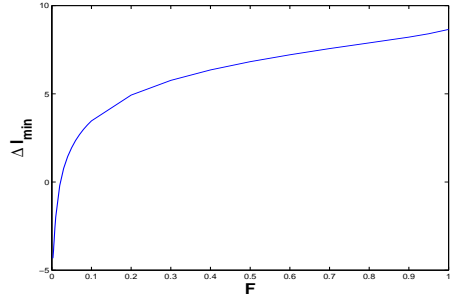}
\end{center}
\caption{The relationship between $\Delta I_{min}$ (unit:bit) and $F$. The parameters are $r=3$ and
$\Sigma^{'2}=100$. }\label{I-F}
\end{figure}

The relationship between $\Delta I_{min}$ and $F$ is useful for detecting eavesdropping. The
analytical expression for these variables is very prolix, so only the numerical solutions, which
are plotted in Fig.\ref{I-F}, are presented. If Eve doesn't exist, i.e., $F=1$, one may easily
obtain the secret information rate $\Delta I_{min}=I(\alpha,\beta)=8.6$ bits. However, the
condition of $F=1$ is too strict in practices. Fortunately, there is an important value $F_{c}$
which may be obtained from Fig.\ref{I-F}, i.e., $F_{c}=0.02$. When $F>F_c$ one has $\Delta
I_{min}>0$, which means Eve's eavesdropping does not influence the security of the final key. While
$F<F_c$ there is a negative information rate. In this case, Alice and Bob has to discard the
communication.

When the line has a transmission $\eta$ over the separation
between Alice and Bob, the best attack strategy for Eve is to take
a fraction $1-\eta$ of the beam and then send the fraction $\eta$
forward through her own lossless line. In this situation, the
eavesdropping can not be detected but the secure QKD is possible
under a proper condition. Eve's maximum information is given by
Eq.(\ref{IAE}) with $\eta_1=\eta_2=\eta$. Numerical calculation
shows $\Delta I\geq 0$ when $\eta\geq 0.728$. Accordingly, Alice
and Bob can perform a deterministic QKD with security when the
line has a transmission $\eta\geq 0.728$. One may recall that the
coherent state quantum key distribution beats the loss limit
$\eta=0.5$ by applying technique of reverse reconciliation
\cite{Grosshans2003} or postselection \cite{Silberhor2002}.
Actually, with the technique of the reverse reconciliation or the
postselection, the loss limit $\eta=0.728$ is anticipated to be
beaten in the proposed scheme \cite{He2005}.
%\textcolor[rgb]{1.00,0.00,0.00}{1. With the technique of the
%reverse reconciliation or the postselection, we anticipate the
%proposed protocol should be still effictive when a substantial
%loss is presented \cite{He2005}.}
%\textcolor[rgb]{1.00,0.00,0.00}{2. Nextly we investigate whether a
%region where $\Delta I=I(\alpha,\beta)-I_{max}(\alpha,\epsilon)>0$
%is entered or not even when a substantial loss is presented by
%applying the reverse reconciliation or postselection.}
%\textcolor[rgb]{1.00,0.00,0.00}{
%With the technique of reverse reconciliation or postselection, a region where $\Delta
%I=I(\alpha,\beta)-I_{max}(\alpha,\epsilon)>0$ is entered even when a substantial loss is presented.

In conclusion, a deterministic QKD scheme using Gaussian-modulated
squeezed states is proposed. The characteristic of no basis
reconciliation yields a two-times efficiency than that of the CV
standard QKD schemes. Especially, the separate control-mode does
not need in the proposed scheme so that the scheme is more
feasible in experimental realization. The fidelity is employed to
detect the eavesdropper and resist the beam splitter attack
strategy. In a lossy channel, a secure scheme requires the
transmission of $\eta>0.728$.

This work is supported by the Natural Science Foundation of China, Grant No. 60472018. GQH
acknowledge fruitful discussions with prof. Z.M.Zhang.


\begin{thebibliography}{}
\bibitem[Gisin(2002)]{Gisin2002}
N. Gisin, G. Ribordy, W. Tittel, and H. Zbinden, Rev. Mod. Phys. \textbf{74,} 145 (2002), and
references therein.

\bibitem[Lo(1999)]{Lo1999}
H. K. Lo and H. F. Chau, Science \textbf{283,} 2050 (1999).

\bibitem[Shor(2000)]{Shor2000}
P. W. Shor and J. Preskill, Phy. Rev. Lett. \textbf{85,} 441 (2000).

\bibitem[Mayers(2001)]{Mayers2001}
D. Mayers, Journal of the ACM \textbf{48,} 351 (2001).

\bibitem[Bostr\"{o}m(2002)]{Bostrom2002}
K. Bostr\"{o}m and T. Felbinger, Phy. Rev. Lett. \textbf{89,} 187902 (2002).

\bibitem[W\'{o}jcikQingYu(2003)]{QingYu2003}
A. W\'{o}jcik, Phy. Rew. Lett. \textbf{90,} 157901 (2003).
%Q. -Y. Cai, Phy. Rew. Lett. \textbf{91,} 109801 (2003).

\bibitem[Lucamarini(2005)]{Lucamarini2005}
M. Lucamarini and S. Mancini, Phy. Rew. Lett. \textbf{94,} 140501 (2005).

\bibitem[Braunstein(2005)]{Braunstein2005}
S. L. Braunstein and P. v. Loock, Rev. Mod. Phy. \textbf{77,} 513 (2005), and reference therein.

\bibitem[Ralph(1999)]{Ralph1999}
T. C. Ralph, Phy. Rev. A \textbf{61,} 010303 (1999).

\bibitem[Hilery(2000)]{Hilery2000}
M. Hilery, Phy. Rev. A \textbf{61,} 022309 (2000).

\bibitem[Gottesman(2001)]{Gottesman2001}
D. Gottesman and J. Preskill, Phy. Rev. A \textbf{63,} 022309 (2001).

\bibitem[Cerf(2001)]{Cerf2001}
N. J. Cerf, M. L\'{e}vy, and G. V. Assche, Phy. Rev. A \textbf{63,} 052311 (2001).

\bibitem[Grosshans(2002)]{Grosshans2002}
F. Grosshans and P. Grangier, Phy. Rev. Lett. \textbf{88,} 057902 (2002).

\bibitem[Silberhorn(2002)]{Silberhorn2002}
Ch. Silberhorn, N. Korolkova, and G. Leuchs, Phy. Rev. Lett. \textbf{88,} 167902 (2002).

\bibitem[Grosshans(2003)]{Grosshans2003}
F. Grosshans et al., Nature(London) \textbf{421,} 238 (2003).

\bibitem[Maurer(1993)]{Maurer1993}
U. M. Maurer, IEEE Trans. Inf. Theory \textbf{39,} 733 (1993).

\bibitem[Shannon(1948)]{Shannon1948}
C. E. Shannon, Bell. Syst. Tech. J. \textbf{27,} 623 (1948).



\bibitem[qoptics(1994)]{Walls1994}
D. F. Walls and G. J. Milburn, \emph{Quantum Optics}. (Springer-Verlag Press, New York, 1997).

\bibitem[Siberhor(2002)]{Silberhor2002}
Ch. Silberhorn, T. C. Ralph, N. L\"{u}tkenhaus, and G. Leuchs,
Phy. Rev. Lett. \textbf{89,} 167901 (2002).

\bibitem[Detail(2005)]{He2005}
Detailed calculations will be presented elsewhere for space
reasons.

\end{thebibliography}
\end{document}